\documentclass{ifacconf}

\usepackage{graphicx}      
\usepackage{natbib}        
\usepackage{blindtext}
\usepackage{amsmath,amsfonts,amssymb}
\usepackage{mathrsfs,mathtools}
\usepackage{enumitem}
\usepackage{xspace}
\usepackage{color}
\usepackage{tikz}
\usepackage{arydshln}

\definecolor{mblue}{rgb}{0,0.4470,0.7410}
\definecolor{morange}{rgb}{0.8500,0.3250,0.0980}
\definecolor{myellow}{rgb}{0.9290,0.6940,0.1250}

\newcommand{\mc}[1]{\mathcal{#1}}

\newcommand{\mr}[1]{\mathrm{#1}}
\newcommand{\mb}[1]{\mathbb{#1}}

\newcommand{\mt}[1]{\mathtt{#1}}
\newcommand{\mbf}[1]{\mathbf{#1}}
\newcommand{\dnx}{n_\mr{x}}
\newcommand{\dny}{n_\mr{y}}
\newcommand{\dnu}{n_\mr{u}}
\newcommand{\dnp}{n_\mr{p}}

\newcommand{\lyap}{Z}

\newtheorem{condition}[thm]{Condition}

\newcommand{\R}{\mb{R}}
\newcommand{\q}{\mr{q}}
\newcommand{\dataset}{\mc{D}_{N_\mr{d}}}

\newcommand{\Dp}{\mc{G}}
\newcommand{\mcG}{\mc{V}}
\newcommand{\Xf}{\overrightarrow{X}}
\newcommand{\Up}{U^\mt{p}}
\newcommand{\Xp}{X^\mt{p}}
\newcommand{\kron}{\otimes}
\newcommand{\ltwo}{\mc{L}_2}

\newcommand{\matlab}{\textsc{Matlab}\xspace}

\newcommand{\EndOfThm}{\hfill$\square$}

\newcommand{\deflen}[2]{%
    \expandafter\newlength\csname #1\endcsname
    \expandafter\setlength\csname #1\endcsname{#2}%
}

\begin{document}
\begin{frontmatter}

\title{Direct data-driven LPV control of nonlinear systems: An experimental result\thanksref{footnoteinfo}} 

\thanks[footnoteinfo]{This work has received funding from the European Research Council (ERC) under the European Union's Horizon 2020 research and innovation programme (grant agreement nr. 714663), the European Space Agency in the scope of the `AI4GNC' project with SENER Aeroespacial S.A. (contract nr. 4000133595/20/NL/CRS), and the Deutsche Forschungsgemeinschaft (DFG, German Research Foundation)-Project No. 419290163. Corresponding author: Chris Verhoek (\texttt{c.verhoek@tue.nl})}

\author[tue]{Chris Verhoek} 
\author[ieem]{Hossam S. Abbas} 
\author[tue,sztaki]{Roland T\'oth}

\address[tue]{Control Systems Group, Dept. of Electrical Engineering, Eindhoven University of Technology, 5600MB Eindhoven, The Netherlands}
\address[ieem]{Institute for Electrical Engineering in Medicine, Universit{\"a}t zu L{\"u}beck, 23558 L{\"u}beck, Germany}
\address[sztaki]{Systems and Control Lab, Institute for Computer Science and Control, 1111 Budapest, Hungary}

\begin{abstract}                
We demonstrate that direct data-driven control of nonlinear systems can be successfully accomplished via a behavioral approach that builds on a \emph{Linear Parameter-Varying} (LPV) system concept. An LPV data-driven representation is used as a surrogate LPV form of the data-driven representation of the original nonlinear system. The LPV data-driven control design that builds on this representation form uses \emph{only} measurement data from the nonlinear system and a priori information on a scheduling map that can lead to an LPV embedding of the nonlinear system behavior. Efficiency of the proposed approach is demonstrated experimentally on a nonlinear unbalanced disc system showing for the first time in the literature that behavioral data-driven methods are capable to stabilize arbitrary forced equilibria of a real-world nonlinear system by the use of only 7 data points.
\vspace{-2mm}
%
%

\end{abstract}


\end{frontmatter}
\deflen{rmwhitebfsec}{-0mm} 
\deflen{rmwhiteafsec}{-3mm} 
\deflen{rmwhitebfssec}{-1.1mm} 
\deflen{rmwhiteafssec}{-3mm} 
\deflen{spacebffigure}{-6mm} 
%
%
\vspace{\rmwhitebfsec}
\section{Introduction}\label{s:introduction}
\vspace{\rmwhiteafsec}
Data-driven analysis and control methods for \emph{Linear Time-Invariant} (LTI) systems that are based on Willems' \emph{Fundamental Lemma} \citep{WillemsRapisardaMarkovskyMoor2005} have become increasingly popular in recent years, as these methods can give guarantees in terms of stability and performance of the closed-loop operation, even if the controller is synthesized from data without any information on the underlying LTI system. Results for LTI systems include, but are not limited to, e.g., data-driven simulation \citep{MarkovskyRapisarda2008}, dissipativity analysis \citep{RomerBerberichKohlerAllgower2019}, and (predictive) control \citep{dePersisTesi2020, CoulsonLygerosDorfler2019}, and many of these methods have seen extensions of their guarantees under the presence of noise. Some extensions of these data-driven methods have been made towards the nonlinear system domain, e.g., \citep{AlsaltiBerberichLopez2021}. However, these results often impose heavy restrictions on the systems in terms of model transformations or linearizations. A promising direction towards data-driven analysis and control of nonlinear systems with guarantees, is the extension of the Fundamental Lemma for \emph{Linear Parameter-Varying} (LPV) systems in \cite{VerhoekTothHaesaertKoch2021}. From this result, extensions on dissipativity analysis \citep{VerhoekBerberich2022}, predictive control \citep{VerhoekAbbasTothHaesaert2021} and state-feedback control \citep{VerhoekTothAbbas2022} have been developed.

The class of LPV systems consists of systems with a linear input-(state)-output relationship, while this relationship itself is varying along a \emph{measurable} time-varying signal -- the \emph{scheduling variable}. The scheduling variable is used to express nonlinearities, time variation, or exogenous effects. This makes the LPV framework highly suitable for nonlinear system analysis and control, by means of using LPV models as \emph{surrogate representations} of nonlinear systems. The LPV framework has shown to be able to capture a relatively large subset of nonlinear systems, by \emph{embedding} the nonlinear system dynamics in an LPV description \citep{Toth2010_book}. Therefore, we show in this work that the data-driven methods for LPV systems are in fact applicable for nonlinear systems, based on the concept of LPV embedding of the underlying nonlinear system.

In the literature, so far data-based control methods of the behavioral kind have been successfully applied in practice on systems that behave fairly linear \citep{MarkovskyDorfler2021Review}. However, successful application of the aforementioned data-driven methods on systems with significant nonlinear behavior has not been accomplished yet, besides in a form of online adaption of an LTI scheme. Therefore, as our primary contribution, we demonstrate in this paper that direct data-driven LPV state-feedback control can achieve tracking of an arbitrary forced equilibria of a real-world nonlinear system. More specifically, we apply the methods in \cite{VerhoekTothAbbas2022} on experimentally obtained measurement data from a nonlinear unbalanced disc system and show that the obtained LPV controller can successfully achieve any angular setpoints and reject disturbances, such as manual tapping of the disk.

In the remainder, we briefly discuss direct data-driven state-feedback controller synthesis for LPV systems in Section~\ref{s:theory}, followed by a description of the experimental results on the unbalanced disc setup in Section~\ref{s:experiment}. The conclusions on the obtained results and recommendations for future research are given in Section~\ref{s:conclusion}.
\vspace{\rmwhitebfsec}
\section{Data-driven LPV state-feedback controller synthesis}\label{s:theory}
\vspace{\rmwhitebfssec}
\subsection{From nonlinear systems to LPV representations}
\begin{figure}
	\centering
	\includegraphics[width=0.8\linewidth]{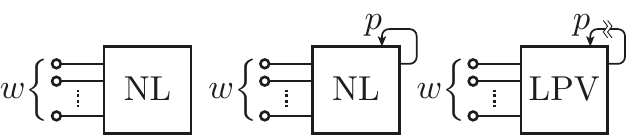}	
	\caption{LPV embedding of a nonlinear system, where $w$ represents the collection of input and output signals.}\label{fig:embedding}
\end{figure}

\vspace{\rmwhiteafssec}
Analysis and control of nonlinear systems using convex approaches can be accomplished by \emph{embedding} the nonlinear system into an LPV representation. Consider a \emph{Discrete-Time} (DT) nonlinear system in \emph{state-space} (SS) form
\begin{subequations}\label{eq:NLsys}
	\begin{align}
		\q x_k&=f(x_k,u_k),\\ 
		y_k &= h(x_k,u_k),
	\end{align}
\end{subequations}
with state variable $x_k\in\mb{R}^{\dnx}$, input variable $u_k\in\mb{R}^{\dnu}$, output variable $y_k\in\mb{R}^{\dny}$, $k\in\mathbb{Z}$ indicating the discrete time-steps with $\q$ representing the forward shift-operator, i.e., $\q x_k=x_{k+1}$, and continuously differentiable functions $f,h$. There exist various methods to embed \eqref{eq:NLsys} into an LPV-SS representation of the form
\begin{subequations}\label{eq:LPVsys}
	\begin{align}
		\q x_k & = A(p_k)x_k + B(p_k)u_k,\label{eq:LPVsys:state}\\ 
		   y_k & = C(p_k)x_k + D(p_k)u_k,
	\end{align}
	where the scheduling variable $p_k\in\mb{P}\subseteq\R^{\dnp}$ is constructed using a so-called \emph{scheduling map} $\psi$, such that
	\begin{equation}
		p_k = \psi(x_k, u_k).
	\end{equation}
\end{subequations}
By assuming that the scheduling $p$ is (i) \emph{measurable}, i.e., it can be calculated from measurable signals via $\psi$, and (ii) allowed to vary independently of the other signals, such as $(x_k,u_k)$ inside a compact, convex set $\mb{P}$ that defines the range of $p$, then we can call \eqref{eq:LPVsys} an embedding of~\eqref{eq:NLsys}. This means that the solution set of~\eqref{eq:NLsys} is embedded in the solution set of \eqref{eq:LPVsys}.
 While linearity is gained by taking these assumptions on $p$, the price to be paid is conservatism of the resulting representation, as the solution set of \eqref{eq:LPVsys} will inevitably contain more solution trajectories than the solution set of \eqref{eq:NLsys}, due to the assumed independence of $p$. The LPV embedding procedure is schematically depicted in Fig.~\ref{fig:embedding}. 
With this embedding strategy, numerous successful applications of LPV control for nonlinear systems have been presented in literature \citep{Toth2010_book, MohammadpourScherer2012}. 

In the sequel, we consider that the scheduling map $\psi$ and the set $\mb{P}$ are known, and with the choice of $\psi$, \eqref{eq:NLsys} can be embedded in the form of \eqref{eq:LPVsys} with the matrix functions $A:\mathbb{P}\rightarrow \mathbb{R}^{n_\mathrm{x}\times n_\mathrm{x}}$ and $B:\mathbb{P}\rightarrow \mathbb{R}^{n_\mathrm{x}\times n_\mathrm{u}}$ and $C=I, D=0$, such that $A$ and $B$ have affine dependence on $p$:
\begin{equation}\label{e:LPVdependency}
	 A(p_k)=A_0+\sum_{i=1}^{n_\mr{p}}p_k^{[i]}A_i, \quad%
	 B(p_k)=B_0+\sum_{i=1}^{n_\mr{p}}p_k^{[i]}B_i,
\end{equation}
 where the coefficients $\{A_i\}_{i=1}^{n_\mr{p}}$ and $\{B_i\}_{i=1}^{n_\mr{p}}$ are real matrices with appropriate dimensions and $p_k^{[i]}$ denotes the $i$\textsuperscript{th} element of the vector $p_k$. Suppose we obtain the data set $\dataset=\{u_k^\mr{d},p_k^\mr{d},x_k^\mr{d}\}_{k=1}^{N_\mr{d}+1}$ from \eqref{eq:LPVsys}. Based on the results of \cite{VerhoekTothAbbas2022}, we now show that we can construct a fully data-driven representation of \eqref{eq:LPVsys},\eqref{e:LPVdependency} using a $\dataset$ that satisfies a \emph{persistence of excitation} condition.

\subsection{Data-driven representations}
\vspace{\rmwhiteafssec}
Considering \eqref{eq:LPVsys:state},~\eqref{e:LPVdependency}, we can separate the coefficients $A_i,B_i$ from the signals as
\begin{equation}\label{e:open-loop-model-based}
	\q x_k = \underbrace{\begin{bsmallmatrix} A_0 & \cdots & A_{n_\mr{p}} \end{bsmallmatrix}}_{\mc{A}} \begin{bmatrix} x_k \\ p_k \kron x_k \end{bmatrix} + \underbrace{\begin{bsmallmatrix} B_0 & \cdots & B_{n_\mr{p}} \end{bsmallmatrix}}_{\mc{B}}\begin{bmatrix} u_k \\ p_k\otimes u_k \end{bmatrix}.
\end{equation}
With $\dataset$, we construct the following matrices that are associated with \eqref{e:open-loop-model-based}:
\begin{subequations}\label{e:data-matrices}\deflen{svspeqfive}{-1.5mm}
\begin{align}
	\hspace{\svspeqfive}U   & = \begin{bmatrix} u_1^\mr{d} & \cdots  & u_{N_\mr{d}}^\mr{d} \end{bmatrix}, \ \Up = \begin{bmatrix} p_1^\mr{d}\otimes u_1^\mr{d} & \cdots &  p_{N_\mr{d}}^\mr{d}\otimes u_{N_\mr{d}}^\mr{d} \end{bmatrix},\\
	\hspace{\svspeqfive}X   & = \begin{bmatrix} x_1^\mr{d}     & \cdots  & x_{N_\mr{d}}^\mr{d} \end{bmatrix}, \ \Xp  = \begin{bmatrix} p_1^\mr{d}\otimes x_1^\mr{d} & \cdots &  p_{N_\mr{d}}^\mr{d}\otimes x_{N_\mr{d}}^\mr{d} \end{bmatrix},
\end{align}
where $U\in\R^{n_\mr{u}\times N_\mr{d}}$, $\Up\in\R^{n_\mr{p}n_\mr{u}\times N_\mr{d}}$, $X\in\R^{n_\mr{x}\times N_\mr{d}}$ and $\Xp\in\R^{n_\mr{p}n_\mr{x}\times N_\mr{d}}$, respectively. Moreover, define $\Xf$ as
\begin{equation}\label{e:data-matrices:Xf}
	\Xf = \begin{bmatrix} x_2^\mr{d} &  \cdots  & x_{N_\mr{d}+1}^\mr{d}\end{bmatrix}\in\R^{\dnx\times N_\mr{d}}.
\end{equation}
\end{subequations}
As \eqref{eq:LPVsys} is linear along $p$, we have 
\begin{equation}\label{e:LPVSysIdent}
	\Xf=\mc{A}
	\begin{bmatrix}
	X\\ \Xp
	\end{bmatrix}+
	\mc{B} \begin{bmatrix}
	U\\ \Up
	\end{bmatrix}.
\end{equation}
Based on this relationship, the LPV system represented by~\eqref{eq:LPVsys:state},~\eqref{e:LPVdependency} can be fully characterized in terms of the data matrices in \eqref{e:data-matrices} as follows
\begin{equation}\label{e:open-loop-data-based}
	\q x_k = \Xf {\underbrace{%
		\begin{bmatrix}	 X \\ \Xp \\ U \\ \Up \end{bmatrix}%
	}_{\Dp}}^\dagger \begin{bmatrix} x_k \\ p_k\otimes x_k \\ u_k \\ p_k\otimes u_k
	\end{bmatrix}.
\end{equation}
The data-based representation \eqref{e:open-loop-data-based} is \emph{well-posed} under the condition that the data set $\dataset$ is \emph{persistently exciting} (PE). The PE condition for $\dataset$, given the representation \eqref{eq:LPVsys},\eqref{e:LPVdependency}, is defined in \cite{VerhoekTothAbbas2022} as follows:
\begin{condition}\hspace{-0pt}\label{cond:rank-cond}
	If $\Dp$ has full \emph{row} rank, i.e., $\mr{rank}\left\{\Dp\right\}=(1+n_\mr{p})(n_\mr{x}+n_\mr{u})$, then $\dataset$ is persistently exciting w.r.t. \eqref{eq:LPVsys}, \eqref{e:LPVdependency} and \eqref{e:open-loop-data-based} is well-defined.
\end{condition}
\begin{rem}
	To satisfy Condition~\ref{cond:rank-cond}, we have that $N_\mr{d} \ge(1+n_\mr{p})(n_\mr{x}+n_\mr{u})$ and, as $\dnx, \dnp, \dnu$, are known it is possible to \emph{explicitly} verify the condition. 
\end{rem}
\subsubsection{Closed-loop data-driven representations:}
Connecting the LPV system with the feedback law $u_k = K(p_k)x_k$, where $K(p_k)$ has affine dependence on $p_k$, i.e., 
\begin{equation}\label{e:controllaw}
	u_k = \big(K_0 + {\textstyle\sum_{i=1}^{\dnp}}p_k^{[i]}K_i\big)x_k = \begin{bsmallmatrix} K_0 & \cdots & K_{\dnp} \end{bsmallmatrix}\begin{bmatrix} x_k\\p_k\otimes x_k \end{bmatrix},
\end{equation}
yields the closed-loop system,
\begin{equation}\label{e:CLLPV-state-feedback-general}
	\q x = \mc{M} \begin{bmatrix}
			x_k \\ p_k\otimes x_k \\ p_k\otimes p_k\otimes x_k
		\end{bmatrix},
\end{equation}
where $\mc{M} = \begin{bsmallmatrix} A_0 + B_0K_0 \, &\,  \bar{A} + B_0\bar{K} + \bar{B}(I_{n_\mr{p}}\otimes K_0) \, &\,  \bar{B}(I_{n_\mr{p}}\otimes \bar{K}) \end{bsmallmatrix}$, with $\bar{A}=\begin{bsmallmatrix} A_1 & \dots & A_{\dnp} \end{bsmallmatrix}$, similarly for $\bar{B},\bar{K}$. With \eqref{e:CLLPV-state-feedback-general}, we can now introduce the following result that provides data-based parametrization of the closed-loop.
\begin{thm}\label{th:closed-loop-data-based-general}\hspace{-0pt}
	Given a $\dataset$ generated by \eqref{eq:LPVsys} that satisfies Condition~\ref{cond:rank-cond}. Let $\Xf$ and $\Dp$ be constructed as in \eqref{e:data-matrices:Xf} and \eqref{e:open-loop-data-based} based on $\dataset$. Then, the closed-loop system \eqref{e:CLLPV-state-feedback-general} is represented equivalently as
	\begin{equation}\label{e:data-based-CLLPV-state-feedback-general}
		\q x_{k}=\Xf \mcG \begin{bmatrix}
		x_k\\p_k\otimes x_k \\ p_k\otimes p_k\otimes x_k
		\end{bmatrix},
	\end{equation} 
	where $\mcG\in\mathbb{R}^{N_\mr{d} \times n_\mathrm{x}(1+n_\mathrm{p}+n_\mathrm{p}^2) }$ is any matrix that satisfies
	\begin{equation}\label{e:consist-cond-general}
		\underbrace{
		\begin{bmatrix}
		I_{n_\mr{x}} & 0& 0\\
		0& I_{n_\mr{p}}\otimes I_{n_\mr{x}}& 0\\
		K_0 & \bar{K} & 0\\
		0 & I_{n_\mr{p}}\otimes K_0 &  I_{n_\mr{p}}\otimes \bar{K} 
		\end{bmatrix}}_{\mathcal{M}_\textsc{CL}}=
		\underbrace{\begin{bmatrix}
		X\\ \Xp \\ U\\ \Up
		\end{bmatrix}}_{\Dp} \mcG,
	\end{equation}
	which we will refer to as the \emph{consistency condition}. \EndOfThm
\end{thm}
\begin{pf}
	See \cite{VerhoekTothAbbas2022}. 
\end{pf}
This result allows to provide direct synthesis methods for the design of data-driven LPV state-feedback controllers that stabilize \eqref{eq:LPVsys}, and hence inherently stabilize the underlying nonlinear system with guaranteed performance.

\vspace{\rmwhitebfssec}
\subsection{Data-driven controller synthesis}
\vspace{\rmwhiteafssec}
\allowdisplaybreaks
With the data-driven representation of the closed-loop LPV system in Theorem.~\ref{th:closed-loop-data-based-general}, LPV state-feedback controller synthesis algorithms can be formulated, which generate controllers using only the information in $\dataset$, while \emph{guaranteeing} stability and performance of the closed-loop. We consider two controller synthesis algorithms, developed in \cite{VerhoekTothAbbas2022}, that yield guarantees in terms of quadratic stability and performance, e.g., $\ltwo$-gain.

Before we can discuss the synthesis methods, we need to introduce a few variables that are necessary to formulate the results. For a $\lyap=\lyap^\top\in\R^{\dnx\times\dnx}$ define 
\begin{subequations}\label{e:vars}
\begin{equation}
	\lyap_0 = \mr{blkdiag}\big(\lyap, \ 0_{\dnx\dnp\times\dnx\dnp} \big),
\end{equation}
and for $\Xf$ in \eqref{e:data-matrices} define
\begin{equation}
	\overrightarrow{\mc{X}} = \mr{blkdiag}\big(\Xf, \ I_{\dnp}\kron\Xf \big).
\end{equation}
Furthermore, for a $\mcG$ satisfying \eqref{e:consist-cond-general}, define the matrices $F_Q$, $\mc{F}$ as 
\begin{multline}\label{e:F_Q}
	\mcG \begin{bsmallmatrix} I_{\dnx} \\ p_k\kron I_{\dnx} \\ p_k\kron p_k\kron I_{\dnx} \end{bsmallmatrix} Z = \\ = \begin{bsmallmatrix} I_{N_\mr{d}} \\ p_k\kron I_{N_\mr{d}} \end{bsmallmatrix}^\top F_Q \begin{bsmallmatrix} I_{\dnx} \\ p_k\kron I_{\dnx}  \end{bsmallmatrix} = \mc{F}\begin{bsmallmatrix} I_{\dnx} \\ p_k\kron I_{\dnx} \\ p_k\kron p_k\kron I_{\dnx} \end{bsmallmatrix}.
\end{multline}
\end{subequations}
With these variables, we can formulate the synthesis method that provides an LPV state-feedback controller that guarantees stability and quadratic performance of the closed-loop system defined by $Q\succeq0,R\succ 0$, in terms of the following theorem.
\begin{thm}\label{th:lqrLPV-result-data-based}\hspace{-0pt}
	Given a $\dataset$ from \eqref{eq:LPVsys} that satisfies Condition~\ref{cond:rank-cond}. There exists an LPV state-feedback controller $K(p)$ in the form of \eqref{e:controllaw} that stabilizes \eqref{eq:LPVsys} and for a given $Q\succeq0,R\succ 0$ minimizes the supremum of 
	\[ J(x,u) = {\textstyle\sum_{k=1}^{\infty}}x_k^\top Q x_k + u_k^\top R u_k, \]
	along all solutions of \eqref{e:CLLPV-state-feedback-general}, if there exist $\lyap=\lyap^\top$, $\Xi$, ${F}_Q$ as in \eqref{e:F_Q}, and $\mc{Y}$, such that
	\begin{subequations}\label{e:synthesis_conditions}
	\begin{gather}
		\left[\begin{array}{c} * \\ \hline  * \end{array}\right]^\top \left[\begin{array}{c|c} \Xi & 0 \\ \hline  0 & W \end{array}\right]^\top	 \left[\begin{array}{c c} L_{11} & L_{12} \\ I & 0 \\ \hline  L_{21} & L_{22} \end{array}\right]\prec 0, \\
		\left[\begin{array}{c} * \\ \hline * \end{array}\right]^\top \underbrace{\begin{bmatrix} \Xi_{11} & \Xi_{12} \\ \Xi_{12}^\top & \Xi_{22} \end{bmatrix}}_{\Xi} \left[\begin{array}{c} I \\ \hline \Delta_p \end{array}\right]   \preceq 0, \quad \Xi_{22} \succ 0,\\
		\begin{bmatrix} \lyap & 0 & 0 \\ 0 & I_{n_\mr{p}}\otimes {\lyap} & 0 \\ Y_0 & \bar{Y}& 0 \\ 0 & I_{n_\mr{p}}\otimes Y_0 &  I_{n_\mr{p}}\otimes \bar{Y} \end{bmatrix} = \Dp \mc{F},
	\end{gather}
	\end{subequations}
	for all $p\in\mb{P}$, where
	\begin{subequations}\label{eq:thm:lqr_vars}
	\begin{align}
		\mc{Y} & =[\,Y_0 \ \ \bar{Y}\,], \\
		\Delta_p & = \mr{blkdiag}\big( p^{[1]}I_{2\dnx},\dots,p^{[n_\mr{p}]}I_{2\dnx}\big), \\
		L_{11} & = 0_{2n_\mr{x}n_\mr{p}\times 2n_\mr{x}n_\mr{p}},\\
		L_{12} & = \begin{bmatrix} \mbf{1}_{n_\mr{p}}\otimes I_{2n_\mr{x}} & 0_{2n_\mr{x}n_\mr{p}\times (n_\mr{x}+n_\mr{u})} \end{bmatrix},\\
		L_{21} & = \begin{bmatrix} 0_{n_\mr{x}\times 2n_\mr{x}n_\mr{p}}\\ I_{n_\mr{p}}\otimes \Gamma_1\\ 0_{n_\mr{x}\times 2n_\mr{x}n_\mr{p}}\\ I_{n_\mr{p}}\otimes \Gamma_2\\ 0_{(n_\mr{x}+n_\mr{u})\times 2n_\mr{x}n_\mr{p}} \end{bmatrix}, \quad \begin{matrix} \Gamma_1 = [ \, I_{\dnx} \ \ 0 \,], \vphantom{\Big)} \\ \Gamma_2 = [\, 0 \ \ I_{\dnx}\,], \vphantom{\Big)}
		\end{matrix}\\
		L_{22} & = \begin{bmatrix} \Gamma_1 & 0 \\ \mbf{1}_{n_\mr{p}}\otimes 0_{n_\mr{x}\times 2n_\mr{x}}&0\\ \Gamma_2 &0\\ \mbf{1}_{n_\mr{p}}\otimes 0_{n_\mr{x}\times 2n_\mr{x}}&0\\ 0 & I_{(n_\mr{x}+n_\mr{u})} \end{bmatrix}, \\
		W & = \begin{bmatrix} Z_0 & F_Q^\top \overrightarrow{\mc{X}}^\top  & \begin{bmatrix} ZQ^{\frac{1}{2}} \\ 0\end{bmatrix} & \mc{Y}^\top R^{\frac{1}{2}} \\ \overrightarrow{\mc{X}}F_Q & Z_0 & 0 & 0 \\ \begin{bmatrix}  Q^{\frac{1}{2}}Z  & 0\end{bmatrix} & 0 & I_{n_\mr{x}} & 0 \\ R^{\frac{1}{2}}\mc{Y} & 0 & 0 &  I_{n_\mr{u}} \end{bmatrix},
	\end{align}
	with $\mbf{1}_{n}=\begin{bsmallmatrix}1 &\cdots& 1\end{bsmallmatrix}^\top\in\mb{R}^n$,
	\end{subequations}
	and $Z_0, \overrightarrow{\mc{X}}, \mc{F}$ as in \eqref{e:vars}. Then, the state-feedback controller $K(p)$ is constructed as
	\begin{equation}\label{e:KSD-stab}
		K_0 = Y_0 \lyap^{-1}, \quad \bar{K} = \bar{Y} (I_{n_\mr{p}}\otimes {\lyap} )^{-1},
	\end{equation}
	where $Z$ is minimizing $\sup_{p\in\mb{P}}\mr{trace}(Z)$ among all possible choices of $Z$ that satisfy \eqref{e:synthesis_conditions}. \EndOfThm
\end{thm}
\begin{pf}
	See \cite{VerhoekTothAbbas2022}. 
\end{pf}
Similarly, we can formulate a data-driven synthesis method that yields an LPV state-feedback controller that guarantees a bound on the $\ltwo$-gain, shaped by matrices $W_\mr{S},W_\mr{R}$ (see \cite{VerhoekTothAbbas2022}),  of the closed-loop system. 
\begin{thm}\label{th:Hinf-LPV-result-data-based}
	Given $W_\mr{S},W_\mr{R}$ and a $\dataset$ generated by \eqref{eq:LPVsys} that satisfies Condition~\ref{cond:rank-cond}. There exists an LPV state-feedback controller $K(p)$ in the form of \eqref{e:controllaw} that stabilizes \eqref{eq:LPVsys} and guarantees that the $\ltwo$-gain of the closed-loop system is less than $\gamma>0$, if there exists a $\lyap=\lyap^\top\succ 0$, a multiplier $\Xi$, ${F}_Q$ as in \eqref{e:F_Q}, and $\mc{Y}$ that satisfy the conditions in \eqref{e:synthesis_conditions} with
	\begin{subequations}
	\begin{align}
		L_{11}& = 0_{2n_\mr{x}n_\mr{p}\times 2n_\mr{x}n_\mr{p}}, \\
		L_{12}&= \begin{bmatrix} \mbf{1}_{n_\mr{p}}\otimes I_{2n_\mr{x}} & 0_{2n_\mr{x}n_\mr{p}\times (2n_\mr{x}+n_\mr{u})} \end{bmatrix},\\
		L_{21}&= \begin{bmatrix} 0_{n_\mr{x}\times 2n_\mr{x}n_\mr{p}}\\ I_{n_\mr{p}}\otimes \Gamma_1\\ 0_{n_\mr{x}\times 2n_\mr{x}n_\mr{p}}\\ I_{n_\mr{p}}\otimes \Gamma_2\\ 0_{(2n_\mr{x}+n_\mr{u})\times 2n_\mr{x}n_\mr{p}} \end{bmatrix}, \\
		L_{22} & =\begin{bmatrix} \Gamma_1 &0\\ \mbf{1}_{n_\mr{p}}\otimes 0_{n_\mr{x}\times 2n_\mr{x}}&0\\ \Gamma_2 &0\\ \mbf{1}_{n_\mr{p}}\otimes 0_{n_\mr{x}\times 2n_\mr{x}}&0\\ 0 & I_{(2n_\mr{x}+n_\mr{u})} \end{bmatrix}, \\
		W & = \begin{bmatrix} Z_0 & (*)^\top & (*)^\top & (*)^\top & 0 \\ \overrightarrow{\mc{X}}F_Q & Z_0 & 0 & 0 & (*)^\top \\ \begin{bmatrix}  W_\mr{S}^{\frac{1}{2}}Z  & 0\end{bmatrix} & 0 & \gamma I_{n_\mr{x}} & 0 & 0 \\ W_\mr{R}^{\frac{1}{2}}\mc{Y} & 0 & 0 &  \gamma I_{n_\mr{u}} & 0 \\ 0 & \begin{bmatrix} I_{n_\mr{x}} & 0 \end{bmatrix} & 0 & 0 & \gamma I_{n_\mr{x}} \end{bmatrix},\label{eq:defW_L2}
	\end{align}
	\end{subequations}
	and with $\mc{Y},\, \Delta_p, \Gamma_1, \Gamma_2$ as in \eqref{eq:thm:lqr_vars}, and $Z_0, \overrightarrow{\mc{X}}, \mc{F}$ as in \eqref{e:vars}. Then, a realization of $K(p)$ is obtained as in \eqref{e:KSD-stab}. \EndOfThm
\end{thm}
\begin{pf}
	See \cite{VerhoekTothAbbas2022}. 
\end{pf}
We can add the minimization of $\gamma$ as an objective while solving \eqref{e:synthesis_conditions} in Theorem~\ref{th:Hinf-LPV-result-data-based} to tighten the upper bound $\gamma$ on the $\ltwo$-gain.
In Section~\ref{s:experiment}, we will compare and apply these methods to synthesize an LPV state-feedback controller for a \emph{nonlinear} system.
\vspace{\rmwhitebfssec}
\subsection{Discussion}
\vspace{\rmwhiteafssec}
We want to again highlight that both these synthesis programs are \emph{only} dependent on (i) the data set $\dataset$ that is measured from the system, and (ii) the assumed knowledge of the scheduling map $\psi$. Furthermore, when the input $u$ is sufficiently exciting, one needs only $(1+\dnp)(\dnx+\dnu)$ data points to be able to synthesize an LPV state-feedback controller for a nonlinear system. It is also important to note that --contrary to well-established results in model-based LPV state-feedback synthesis \citep{rugh2000research, rotondo2014robust}-- our synthesis methods allow to have a scheduling dependent $B$ matrix in the LPV representation \eqref{eq:LPVsys} of the true underlying system, although such a representation/model is never computed in our methodology. Finally, the data that is used in the synthesis methods of Theorems~\ref{th:lqrLPV-result-data-based} and \ref{th:Hinf-LPV-result-data-based} is assumed to be noise-free, just as in the initial results for LTI systems \citep{MarkovskyRapisarda2008}. Extending our results to handle noise-infected data is an important future research direction. The role of regularization, as discussed in the recent paper \cite{breschi2022role}, is expected to be a promising approach to further generalize this method for noisy data.

\vspace{\rmwhitebfsec}
\section{Experimental setup}\label{s:experiment}
\vspace{\rmwhiteafsec}
We will now apply the discussed data-driven LPV state-feedback synthesis methods to design a controller for an experimental nonlinear unbalanced disc setup and implement the designed controller on it.

\vspace{\rmwhitebfssec}
\subsection{Setup description}
\begin{figure}
	\centering
	\includegraphics[height=30mm]{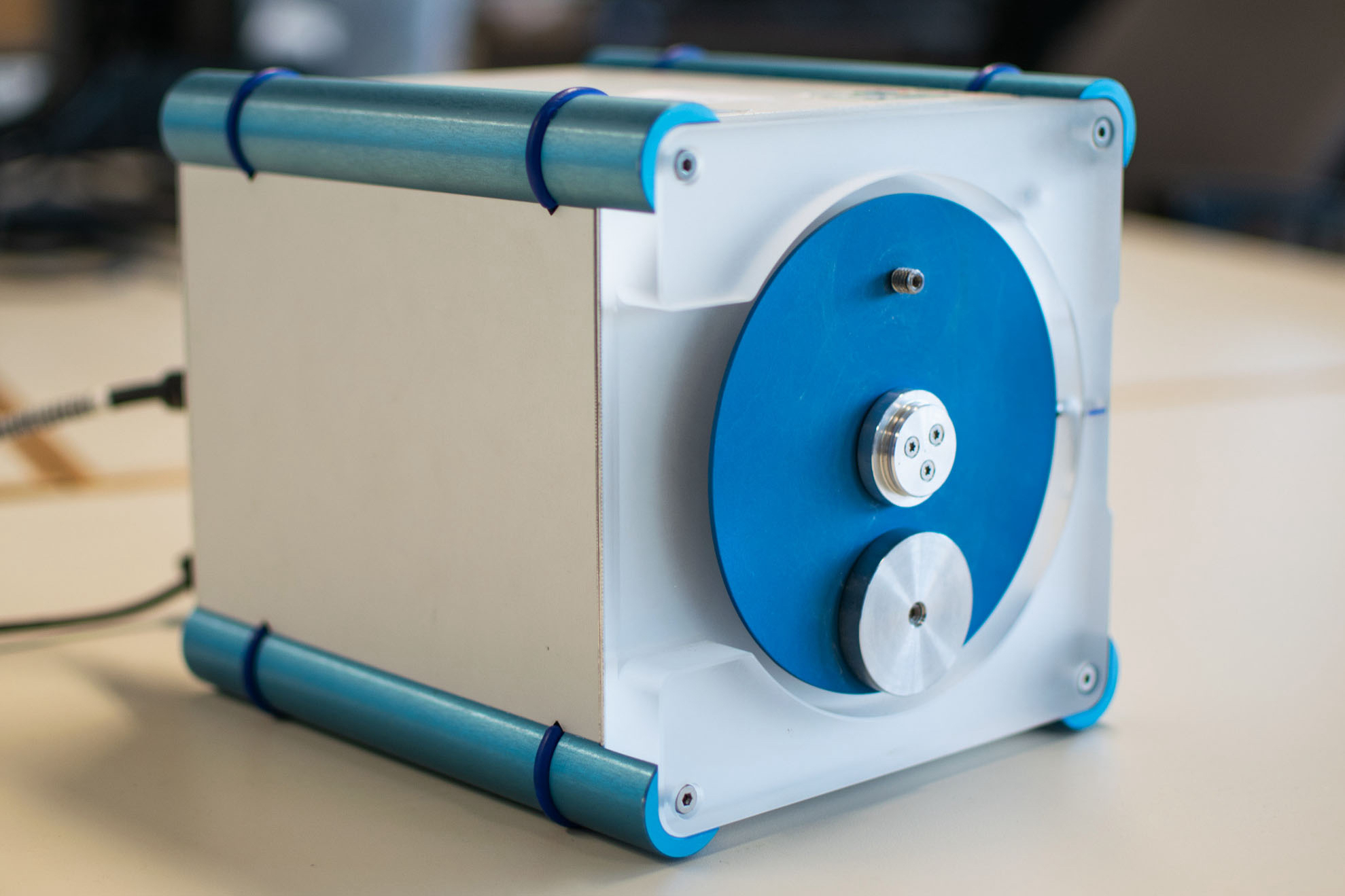}
	\vspace{-1mm}
	\caption{The unbalanced disc experimental setup.}
	\label{fig:unbaldisc}
\end{figure}

\vspace{\rmwhiteafssec}
The experimental setup that we consider is an unbalanced disc system, depicted in Fig.~\ref{fig:unbaldisc}. 
The unbalanced disc setup consists of a DC motor connected to a disc with an off-centered mass, and hence it mimics the behavior of a rotational pendulum. The angular position of the disc is measured using an incremental encoder.
The continuous-time nonlinear dynamics of the unbalanced disc system are represented by the following ordinary differential equation,
\begin{equation}\label{e:unbalanced-disc}
\ddot{\theta}(t)=-\frac{mgl}{J}\sin(\theta(t))-\frac{1}{\tau}\dot{\theta}(t)+\frac{K_\mr{m}}{\tau}u(t),
\end{equation}
where $\theta$ is the angular position of the disc, $u$ is the input voltage to the system, which is its control input, and $m,g,l,J,\tau,K_\mr{m}$ are the physical parameters of the system. In this setup, $u$ is applied with a zero-order-hold actuation and is saturated at $\pm10$~[V]. 
In \cite{KulcsarDongWingerdenVerhaegen2009, KoelewijnToth2019_unbalanced}, the physical parameters for this setup are estimated using LPV identification methods, and model-based LPV controllers have been successfully applied to it using the estimated parameters \citep{AbbasEtAl2017_multipathfblin}. We actuate and measure the system with synchronized sampling at a sampling time of $T_\mr{s}=0.01$ seconds.  Embedding \eqref{e:unbalanced-disc} into an LPV representation can be established by defining the scheduling as $p(t) = \mr{sinc}(\theta(t))=\tfrac{\sin(\theta(t))}{\theta(t)}$, where $\psi(\theta)=\mr{sinc}(\theta(t))$ is the {scheduling map}. Hence, $\mb{P}$ is considered as the \emph{range} of the $\mr{sinc}$ function, i.e., $\mb{P}=[-0.22,\; 1]$. By choosing $x = [\theta \  \dot\theta]^\top$, we can write \eqref{e:unbalanced-disc} as a continuous-time LPV-SS representation:
\begin{subequations}\label{eq:unbal_LPVSS}
	\begin{align}
		\dot x (t) & = \begin{bmatrix} 0 & 1 \\ -\tfrac{mgl}{J}p(t) & -\tfrac{1}{\tau} \end{bmatrix} x(t) + \begin{bmatrix} 0 \\ \tfrac{K_\mr{m}}{\tau} \end{bmatrix} u(t), \\
		y(t) & = x(t).
	\end{align}
\end{subequations}
It is clear that \eqref{eq:unbal_LPVSS} is affinely dependent on $p(t)$. 
Moreover, $p$ can be obtained \emph{through} the aforementioned scheduling map by measuring $\theta$ with the encoder. Therefore, we can use our data-driven synthesis methods, discussed in Section~\ref{s:theory}, for the design of a DT controller for this system without knowing at all the DT equivalent of \eqref{e:unbalanced-disc} under the considered $T_\mr{s}$, the parameter values, or \eqref{eq:unbal_LPVSS}. We only assume that $\psi$, i.e., the calculated sequence of $p_k$, is available.

\vspace{\rmwhitebfssec}
\subsection{Data generation}
\vspace{\rmwhiteafssec}
The system is controlled using a real-time \matlab environment on a MacBook Pro (2020), which communicates with the setup through an USB port. More details on the \matlab environment can be found in \cite{KulcsarDongWingerdenVerhaegen2009}. The \matlab environment ensures synchronous actuation and measurement of the setup. We obtain the output of the incremental encoder ($\theta_k$) and an estimate of the angular speed $\dot\theta_k$ as measurements.

We obtain our data-dictionary $\dataset$ by exciting the system with a randomly generated input in the range $[-10,\,10]$, filtered through a low-pass filter to restrict the frequency range of the excitation to the normal operation range of the setup. Condition~\ref{cond:rank-cond} provides that  $N_\mr{d}\ge 6$, i.e., \mbox{$T_\mr{exp}\ge 60$ [ms].} To account for noise induced by the encoder and the estimation error of $\dot\theta_k$, we gathered 1 second worth of data from which we selected our data-dictionary that is used in the synthesis algorithms.

\vspace{\rmwhitebfssec}
\subsection{Controller synthesis}
\vspace{\rmwhiteafssec}
For this experimental demonstration, we considered the design of two controllers:
\begin{enumerate}[label={Controller \arabic*: }, align=left, ref={Controller~\arabic*}]
	\item LPV state-feedback controller guaranteeing stability and optimal quadratic performance (Th.~\ref{th:lqrLPV-result-data-based}). \label{controller1}
	\item LPV state-feedback controller guaranteeing stability and a minimal $\ltwo$-gain of the closed-loop system given $W_\mr{S},W_\mr{R}$ according to Th.~\ref{th:Hinf-LPV-result-data-based}.\label{controller2}
\end{enumerate}

\begin{figure}
	\centering
	\includegraphics[scale=1, trim=0mm 0.95mm 0mm 0.7mm, clip]{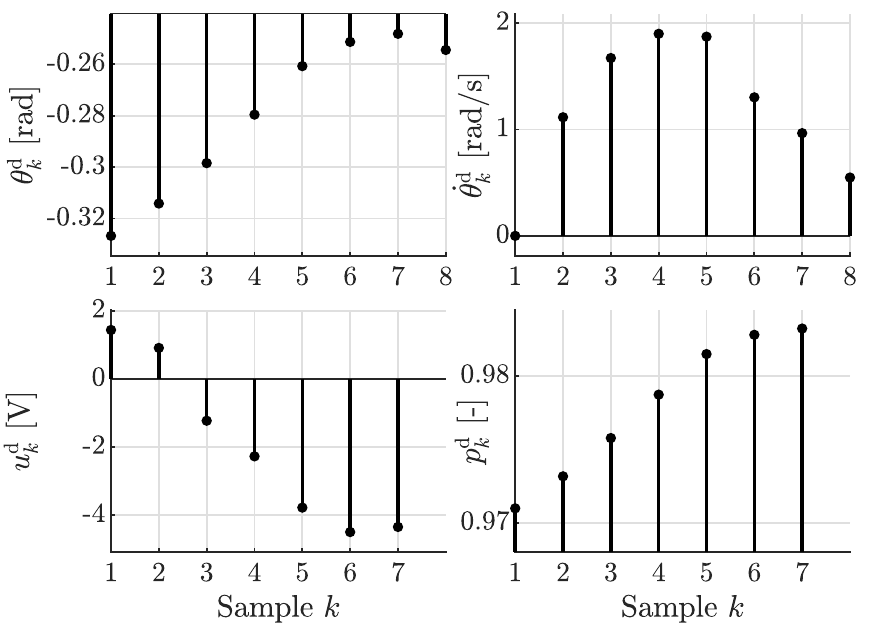}
	\vspace{\spacebffigure}
	\caption{Data-dictionary used for controller synthesis with $N_\mr{d}=7$.}
	\label{fig:datadic}
\end{figure}

From the obtained measurement data, we randomly select a window of $N_\mr{d}=7$ samples, which satisfies Condition~\ref{cond:rank-cond} to create our data-dictionary $\dataset$. Our data-dictionary is depicted in Fig.~\ref{fig:datadic}. As $\dataset$ satisfies Condition~\ref{cond:rank-cond}, we will use it to represent the behavior of the nonlinear unbalanced disc system. 
Note that the scheduling signal $p_k$ is calculated by propagating the measurement of $\theta_k$ through the scheduling map. From the data-dictionary, we construct the matrices $X,U,\Xp,\Up,\Xf$ according to \eqref{e:data-matrices}, and the data-driven representation $\Dp$, which are used in the synthesis procedures of Theorems~\ref{th:lqrLPV-result-data-based}~and~\ref{th:Hinf-LPV-result-data-based}.

We are now ready to synthesize \ref{controller1} and \ref{controller2}. In order to improve numerical conditioning, we re-scale and re-center the scheduling in the synthesis problems such that $\tilde{\mb{P}}=[-1,\;1]$.  Solving the synthesis problem for \ref{controller1} with $Q=\mr{diag}(4,\;0.1)$, $R=3.5$ in \matlab, using YALMIP with solver MOSEK, yields
\begin{align*}
	K_{\mr{QP},0} &= \begin{bmatrix} -8.7239 & -1.0733 \end{bmatrix} \\
	K_{\mr{QP},1} &= \begin{bmatrix} -0.4894 & 0.0021 \end{bmatrix}.
\end{align*}
For numerical conditioning, we add a regularization in the cost function for the synthesis of \ref{controller2} in terms of $\min~\gamma+\lambda\mr{trace}(\lyap)$.
Synthesizing \ref{controller2} with $W_\mr{S}=\mr{diag}(1.5\cdot10^{-2},\;2\cdot10^{-5})$, $W_\mr{R}=3.06\cdot10^{-3}$ and $\lambda=10^{-6}$ yields $\gamma=3.3$ with
\begin{align*}
	K_{\ltwo,0} &= \begin{bmatrix} -15.1412 &  -1.8074 \end{bmatrix} \\
	K_{\ltwo,1} &= \begin{bmatrix} 8.9793 & 1.0102 \end{bmatrix}.
\end{align*}
Just like in model-based design, we have chosen the values for $Q, R, W_\mr{S}$ and $W_\mr{R}$ by iterative tuning. We want to emphasize that we have been able to synthesize these LPV state-feedback controllers using \emph{only} the 7 data points shown in Fig.~\ref{fig:datadic}. We will now implement these controllers on the experimental setup.

\vspace{\rmwhitebfssec}
\subsection{Experiments}
\vspace{\rmwhiteafssec}
For the experimental verification of our synthesized controllers, we consider two scenarios; a disturbance rejection scenario and a reference tracking scenario.
%
\subsubsection{Disturbance rejection:} 
In this scenario, the controller will regulate the states of the unbalanced disc at the origin, while we disturb the system by displacing the disc by hand. We define the origin of the unbalanced disc system as the upright position. For this scenario, we only present results with \ref{controller1}, as it was not possible to apply exactly the same disturbance for both experiments. We ran the controller online for 20 seconds, and manually disturbed the system 5 times by pulling the mass away from the upright position. 
\begin{figure}
	\centering
	\includegraphics[scale=1, trim=0mm 0.95mm 0mm 0.7mm, clip]{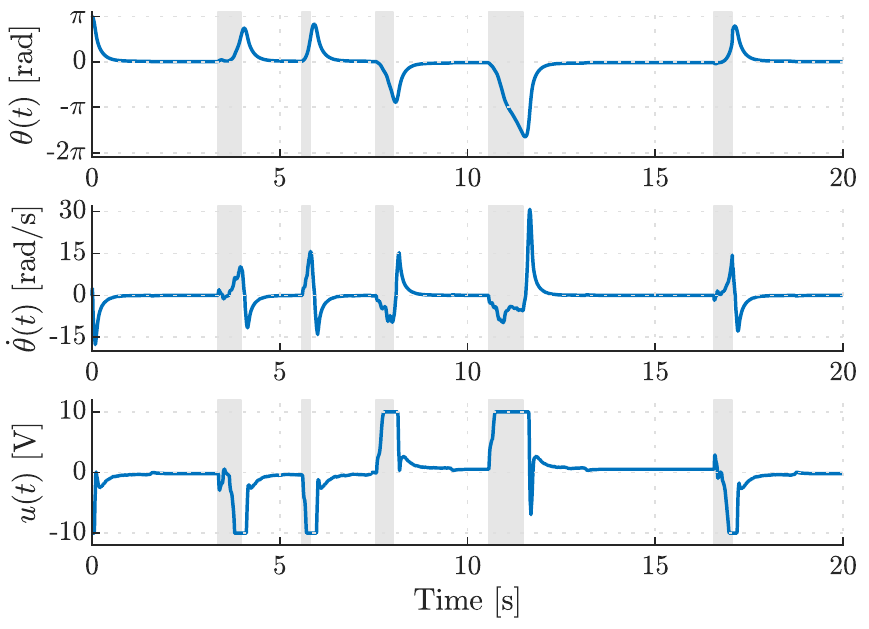}
	\vspace{\spacebffigure}
	\caption{Experimental results with \ref{controller1} in a disturbance rejection scenario. The gray areas indicate when we disturbed the experimental setup by hand.}
	\label{fig:disturbance}
\end{figure}
Fig.~\ref{fig:disturbance} shows the obtained measurement results during the experiments. The gray-shaded areas in the plots indicate the instances where we have manually disturbed the system. A video of the experiment is available at \texttt{https://youtu.be/m9l61lR1Fw4}. From the figure, we can conclude that the designed LPV state-feedback controller, which is synthesized using only 7 time measurements, can robustly regulate the \emph{nonlinear} system at the setpoint of 0 [rad] for the full operating range.
%
\subsubsection{Reference tracking:}
\begin{figure}
	\centering
	\includegraphics[scale=1, trim=0mm 0.95mm 0mm 0.7mm, clip]{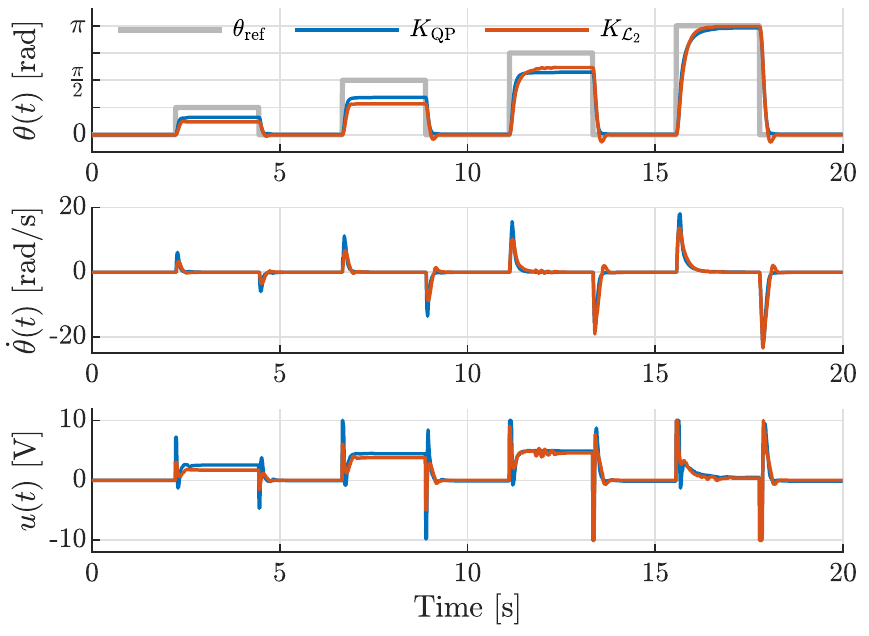}
	\vspace{\spacebffigure}
	\caption{Experimental results with changing setpoint.}
	\label{fig:reference}
\end{figure}

In this scenario, we asses if the designed controllers can realize arbitrary setpoints for the nonlinear system. We select the sequence of setpoints as $\theta_\mr{ref}=\{0,\,\tfrac{\pi}{4},\,0,\,\tfrac{\pi}{2},\,0,\,\tfrac{3\pi}{4},\,0,\,\pi,\,0\}$ that $\theta(t)$ should track in the experiment. In order to ensure reference tracking, we deploy the controllers in terms of
\begin{equation}\label{e:mod_contr_law}
	u_k = (K_0 + K_1p_k)\left(\begin{bmatrix}\theta_k \\ \dot{\theta}_k \end{bmatrix}-\begin{bmatrix} \theta_{k,\mr{ref}} \\ 0 \end{bmatrix} \right).
\end{equation}
Note that we do not apply a feedforward term here, i.e., the controller considers $(x,u)=(\begin{bmatrix} \theta_\mr{ref} & 0 \end{bmatrix}^\top, 0)$ to be equilibrium points. 
Running the experiment for 20 seconds with either \ref{controller1} or \ref{controller2} interconnected with the setup yields the measurements plotted in Fig.~\ref{fig:reference}. A video of the experiments is available at \texttt{https://youtu.be/SyyUVy1sPsc}.
The results show that the closed-loop system achieves reference tracking along the full operating range of the nonlinear system with an LPV controller that is designed using only 7 data-points recorded from the system. Note however that due to the lack of integral action or a feedforward term, there is a small steady-state error for some reference points, which is in accordance with the theory of standard state-feedback design.
%
\subsection{Discussion}
\vspace{\rmwhiteafssec}
We have shown that the designed LPV controllers can achieve the stability and performance objectives on the nonlinear system. This is accomplished by measuring a data-dictionary of input-scheduling-state data, where the scheduling is constructed using a given scheduling map. Although, the current results assume the data is noise free, as was the case in the initial results on LTI system, our experimental results show that there is a certain level of robustness of the data-driven control laws that are obtained via this design procedure. However, proper guarantees of performance and stability of the closed-loop are not available in case of noise-infected data. We want to note that the synthesis algorithms gave numerical problems for some of the selected windows from the original 1-second-long data set. Hence, even though persistency of excitation holds in terms of a rank condition, it does matter that what the conditioning number is associated with `$\Dp$', i.e., the amount of relative information present in the chosen window. A more quantitative measure of information content in the data could certainly be useful to avoid such numerical difficulties. 

\vspace{\rmwhitebfsec}
\section{Conclusions and outlook}\label{s:conclusion}
\vspace{\rmwhiteafsec}
The experimental results presented in this work demonstrate that direct data-driven control of nonlinear systems can be established via LPV data-driven methods. Furthermore, for the deployed LPV data-driven control design for the unbalanced disc system, only 7 data-points are required to ensure stability and performance on the setup. A next step for further improvement of the proposed methodology is to make it robust to noisy data.

\bibliography{refs_ifac2023}

\end{document}